\documentclass[
 reprint,
superscriptaddress,
aip,
jcp,
 amsmath,amssymb,
floatfix,
]{revtex4-1}
\usepackage{dcolumn}
\usepackage{bm}
\usepackage{geometry}
\usepackage{graphicx} 

\usepackage{float}
\usepackage{multirow}
\usepackage{booktabs} 
\usepackage{array} 
\usepackage{paralist} 
\usepackage{verbatim} 
\usepackage{subfig} 
\usepackage{rotating}

\usepackage{fancyhdr} 
\usepackage{threeparttablex}
\usepackage{amsmath, amsthm}
\usepackage{afterpage}
\usepackage[nottoc,notlof,notlot]{tocbibind} 
\usepackage[titles,subfigure]{tocloft} 

\graphicspath{{./}}
\DeclareGraphicsExtensions{.eps}

\begin{document}

\title{Decomposition of Unitary Matrices for Finding Quantum Circuits: Application to Molecular Hamiltonians}
\author{Anmer~Daskin}\affiliation{Department of Computer Science, Purdue University, West Lafayette, IN, 47907 USA}
\author{Sabre~Kais}\email[Corresponding author. Email: ]{kais@purdue.edu}
\affiliation{Department of Chemistry, Physics, and Birck Nanotechnology Center,Purdue University,
West Lafayette, IN 47907 USA} 

\begin{abstract}
 Constructing appropriate unitary matrix operators for  new quantum algorithms and finding the minimum cost
gate sequences for the implementation of these unitary operators is of fundamental importance in the field
of quantum information and quantum computation. Evolution of quantum circuits faces two major challenges:
 complex and huge search space and the high costs of simulating quantum circuits on 
classical computers. Here, we use the group leaders optimization algorithm
to decompose a given unitary matrix into a proper-minimum cost quantum gate sequence. We test the method on the known decompositions of Toffoli gate, 
the amplification step of the Grover 
search algorithm, the quantum Fourier transform, and
the sender part of the quantum teleportation. Using this procedure, we present the circuit designs for the simulation of the unitary propagators of the
  Hamiltonians for the hydrogen and the water molecules.  
The approach is general and can be applied to generate the sequence of quantum gates for larger molecular systems.

\end{abstract}

\maketitle
\section{Introduction}

Quantum computation  promises to solve fundamental, yet otherwise intractable problems
in many different fields.  
 To advance 
 the quantum computing field, finding  circuit designs to execute algorithms on quantum computers (in the circuit 
 model of quantum computing) is important.
Therefore, it is of fundamental importance to develop new methods with which to overcome
the difficulty in
 forming a unitary matrix describing the algorithm (or the part of the computation),
and the difficulty to decompose this matrix into the known quantum gates \cite{citeulike:541803}.  Realizing the theoretical problems of quantum computers 
requires the overcoming decoherence problem \cite{Whaley}. Recently, West et al. demonstrate numerically 
that high fidelity quantum gates are possible in the frame work of quantum dynamic and decoupling \cite{West}.

It has been shown that the ground and excited state energies  of small molecules
can be carried out on a quantum computer simulator using a recursive phase-estimation 
algorithm \cite{Abrams,Alan-Science,Hefeng,PhysRevE.59.2429}. Lanyon et al. reported the application of photonic 
quantum computer technology to calculate properties of the hydrogen molecule in a minimal
basis \cite{Lanyon}. For the simulation of quantum systems, it is needed to find efficient quantum circuits. 

The problem in the  decomposition of a given unitary 
matrix  into a sequence of quantum logic gates can be presented as an  
optimization problem. Williams and Gray  \cite{Williams99automateddesign} suggested the use 
of genetic programming
 technique to find new circuit designs for  known algorithms, and also presented results for  the
quantum teleportation. 
Yabuki, Iba \cite{Yabuki00geneticalgorithms} and Peng et al. \cite{10.1109/ICCIT.2009.178} 
focused on circuit designs for the quantum teleportation 
by using different genetic algorithm techniques. Spector \cite{1211041} explains the use of 
genetic programming to explore new quantum algorithms. 
Stadelhofer  \cite{CambridgeJournals:1903220} used the genetic algorithms to evolve black box quantum algorithms.
 There are also some other works \cite{oaieldorado0x0007ce5c, 787893, DBLPconfgeccoMasseyCS04} 
which evolve quantum algorithms 
or circuits by using the genetic programming or the genetic algorithms. 
Review of these procedures can  be found 
in \cite{DBLPjournalsgpemGeppS09}.

The evolution of quantum circuits faces two major challenges:
 complex and huge search space and the high costs of simulating quantum circuits on 
classical computers. In this paper,  
we use our recently developed group leaders optimization algorithm (GLOA)\cite{Daskin}-an evolutionary algorithm-to decompose the unitary matrices
 into a set of quantum gates. 
We show how our approach can be used to find the circuit representation of a quantum algorithm or the unitary propagator of a quantum system which is
essential to perform the simulation. The approach was tested on the operators of the Grover search algorithm; the sender part of the quantum teleportation; 
the Toffoli gate; and the quantum Fourier transform. It was also used to find the circuit designs for the simulations of the hydrogen and the water molecules on quantum computers.

This paper is organized as follows: after giving the essentials of the objective function in the next section, 
we give the optimization results for quantum algorithms in Sec.\ref{Sec:Algorithms}. And in Sec.\ref{Sec:Hamiltonians} we 
explain how the method can be used to design circuits for the simulation of molecular Hamiltonians and present the circuit designs and  their simulations within the phase estimation algorithm  for the hydrogen and water molecules. 

\section{The Overview of the Optimization Scheme}

\subsection{The Objective}
The objective of the optimization process is to find quantum circuits with minimum costs and errors.
Thus, there are two factors which need to be optimized within the optimization: the error and the cost of the circuit. 
 The minimization of the error to an acceptable level is more important than the cost in order to get more accurate 
and reliable results in the optimization process (The importance of the cost 
and the error in the approximated circuit can be adjusted by an objective function constituting both with some weights);  
hence, in the optimization the circuit giving a better approximation to the solution is always preferred over the other circuit with a lower cost.

\subsubsection{The Fidelity Error}
\label{section2}
For the unitary matrix representation of a candidate approximation circuit, $U_a$, in the optimization and  a given target unitary matrix $U_t$, 
William et al. \cite{Williams99automateddesign} described the quality of the circuit as follows:
\begin{equation}
\label{Williamserror}
f(U_a,U_t)=\sum_{i=1}^{2^n}\sum_{j=1}^{2^n}{|U_{t(ij)}-U_{a(ij)}|}, 
\end{equation}
where $U_t$ and $ U_a \in U^{2^n}$ and $f=0$ when $U_t=U_a$.
Since the global phase differences between two quantum systems are physically indistinguishable; when $U_t$ and $U_a$ are different only in terms of  their global phases, 
the value of $f$ should be zero. However, Eq.\ref{Williamserror} is
unlikely to produce zero for this case.   
Here, instead of Eq.\ref{Williamserror}, we use the trace fidelity error which ignores the global phase differences; hence, makes the optimization easier by diversifying  
the reachable solutions for a problem in complex space. The trace fidelity is given by:
\begin{equation}
\label{correctness1}
\mathcal{F}=\frac{1}{N}\left|Tr\left(U_aU_t^\dagger\right)\right|,
\end{equation}
where $N=2^n$ ($n$ is the number of qubits); the symbol $\dagger$ represents the complex 
conjugate transpose of a matrix; and $Tr(..)$ is the trace of a matrix. 
Since the product of two unitary matrices is another unitary matrix all eigenvalues of which have absolute value 1,
$\mathcal{F}$ is always in the range $\left[0,1\right]$ and is equal to 1 when $U_a=U_t$.  The fidelity error  used 
in the optimization to measure how similar the unitary operators $U_a$ and $U_t$ are is defined as \cite{Vidal}:
\begin{equation}
 \epsilon=1-\mathcal{F}^2,
\end{equation}
where $\mathcal{F}$ is squared to increase the effects of small fidelity changes in the error.

\subsubsection{The Cost of a Circuit}
The cost of a circuit describes the level of ease with which this circuit is implemented; 
in order to  make the implementation of a circuit easier and the circuit less error-prone,  
the cost  also needs to be optimized by minimizing the number of gates in the circuit. 

However, defining the cost of a circuit is not an easy task due to the fact that each quantum computer model may have a different cost for a given quantum gate.
Here,  the costs of a one-qubit gate and a control (two-qubit) gate are  defined as 1 and 2 respectively. Since the implementation of a multi-control gate (n-qubit network) requires $\Theta(2^n)$ (see Ref.\cite{Barenco}),  its cost is defined as $2^n$ where $n$ is the number of qubits on which the gate is operating. The cost of a circuit is found by summing up the costs of the gates composing the circuit. For instance, 
the cost of the circuit in Fig.\ref{fig:toffoli} is 10.
The real implementation cost of a circuit may be different than this abstraction; nevertheless, 
a lower-cost circuit found in the optimization likely costs less than a higher-cost circuit 
in the implementation.

\subsection{The Representation of Quantum Gates in the Optimization}

\begin{figure}[htbp]%
\centering
\includegraphics[width=2.5in]{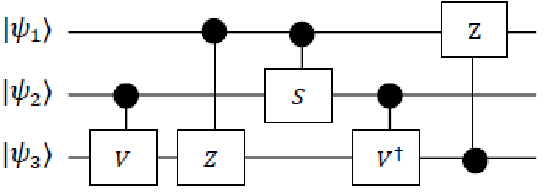}
\caption{The circuit design for the Toffoli gate.}
\label{fig:toffoli}
\end{figure}

We use  similar representation method to the method of the cartesian genetic programming \cite{1389075} in which each function set and inputs
 (in our case, gates and qubits) are represented as integers and genotypes including the inputs and the gates are represented as integer strings.
  The difference is: since the quantum gates may have an effect on the whole system, 
the gates should be represented in time steps; that means we cannot give the same inputs to two different gates at the same time as it is done in classical circuits. 
 In the  strings of the genotypes each four numbers represent a gate, the qubit on which the gate operates, the control qubit, and 
the angle for the rotation gates. The integers for  gates are determined by looking at the indices of the gates in the gate set. 
For a given set of gates $\{V, Z, S, V^\dagger\}$ (see Appendix for the matrix representation of these gates), an example numeric string  representing the circuit in Fig.\ref{fig:toffoli} is as:

\textbf{1} 3 2 0.0; \textbf{2} 3 1 0.0; \textbf{3} 2 1 0.0; \textbf{4} 3 2 0.0; \textbf{2} 1 3 0.0,\\
where the each group of four numbers separated by semicolons describes the each quantum gate in the circuit: the first numbers with bold fonts identify the gates ($\{V=1, Z=2, S=3, V^\dagger=4\}$),
 the last numbers are the values of the angles between 0 and $2\pi$ (for non-rotation gates it is considered 0.), and the middle integers are the target and the control qubits, respectively, (the semicolons and the bold fonts do not appear in the 
real implementation). For  multi-control gates the qubits between the control and the target qubits are also considered as control qubits.

The length of a numeric string is the maximum number of gates a circuit can include. The required maximum number of gates can be very large:
 Suppose $U$ acts on a $2^n$-$dimensional$ Hilbert subspace. Then $U$ may be written as a product of at most 
$2^{n-1}(2^{n}-1)$ two-level unitary systems  \cite{citeulike:541803}.
 For 5 qubits, there may be 496 two-level unitary matrices required to form a given $U$. In our optimization cases,
the maximum number of gates ($max_{gates}$) is limited to 20 gates.  
\subsection{The Group Leaders Optimization Algorithm}

The group leaders optimization algorithm (GLOA) described in more detail in \cite{Daskin} is a simple  and effective global optimization algorithm 
that models the influence of leaders in social groups as an optimization tool.
 The general structure of the algorithm is made up by dividing the population into several disjunct groups
 each of which has its leader (the best member of the group) and members.
 The algorithm which is different from the earlier evolutionary algorithms and the pivot method algorithm 
\cite{PhysRevE.55.1162,Nigra1999433,serra:7170} consists of two parts. In the first part, the member itself-the group leader with possible
 random part-and a new-created random solution are used to form a new member. 
 This mutation is defined as: 
\begin{equation}
\begin{split}
 new\ member &= r_1\ portion\ of\ old\ member\\ & \cup\ 
r_2\ portion\ of\ leader \\ &\cup\ r_3\ portion\ of\ random,
\end{split}
\end{equation}
where $r_1$, $r_2$, and $r_3$ determine the rates of the old member, the group
 leader, and the new-created random solution into the new formed member 
and they sum to 1. In our case, they are set as $r_1=0.8$ and $r_2=r_3=0.1$.  The mutation for the values of the all angles in a numerical string is done according to the arithmetic expression: $angle_{new}=r_1\times angle_{old}+r_2\times angle_{leader}+r_3\times angle_{random}$, where  $angle_{old}$, the current value of an angle, is  mutated:  $angle_{new}$ , the new value of the angle, is formed by combining a random value and the corresponding leader of the group of the angle and the current value of the angle with the coefficients $r_1$, $r_2$, and $r_3$. The mutation for the rest of the elements in the string means the replacement of its elements  by the corresponding  elements of the leader  and a newly generated random string with the rates $r_2$ and $r_3$.

 In addition to the mutation, in each iteration for each group of the population 
one-way-crossover (also called the parameter transfer) is done  between a chosen random member from 
the group and a random member from a different-random group. This operation is  
mainly replacing some random part  of a member with the equivalent part of a random member
 from a different group. The amount of the transfer operation for each group is defined by a parameter called transfer rate, here, which is defined as: 
$\frac{4\times max_{gates}}{2}-1$, where the numerator is the number of variables forming a numeric string in the optimization. 

\textbf{The replacement criteria}: If a new formed (or mutated) member 
gives less error-prone solution to the problem than the corresponding old member, or they have the same error values but the cost of the new member is less than 
the old member, 
then the new member survives and replaces the old member; otherwise, the old member remains for the next iteration and the new formed one is discarded.  
 
\begin{figure*}[htbp]
\includegraphics[width=5in]{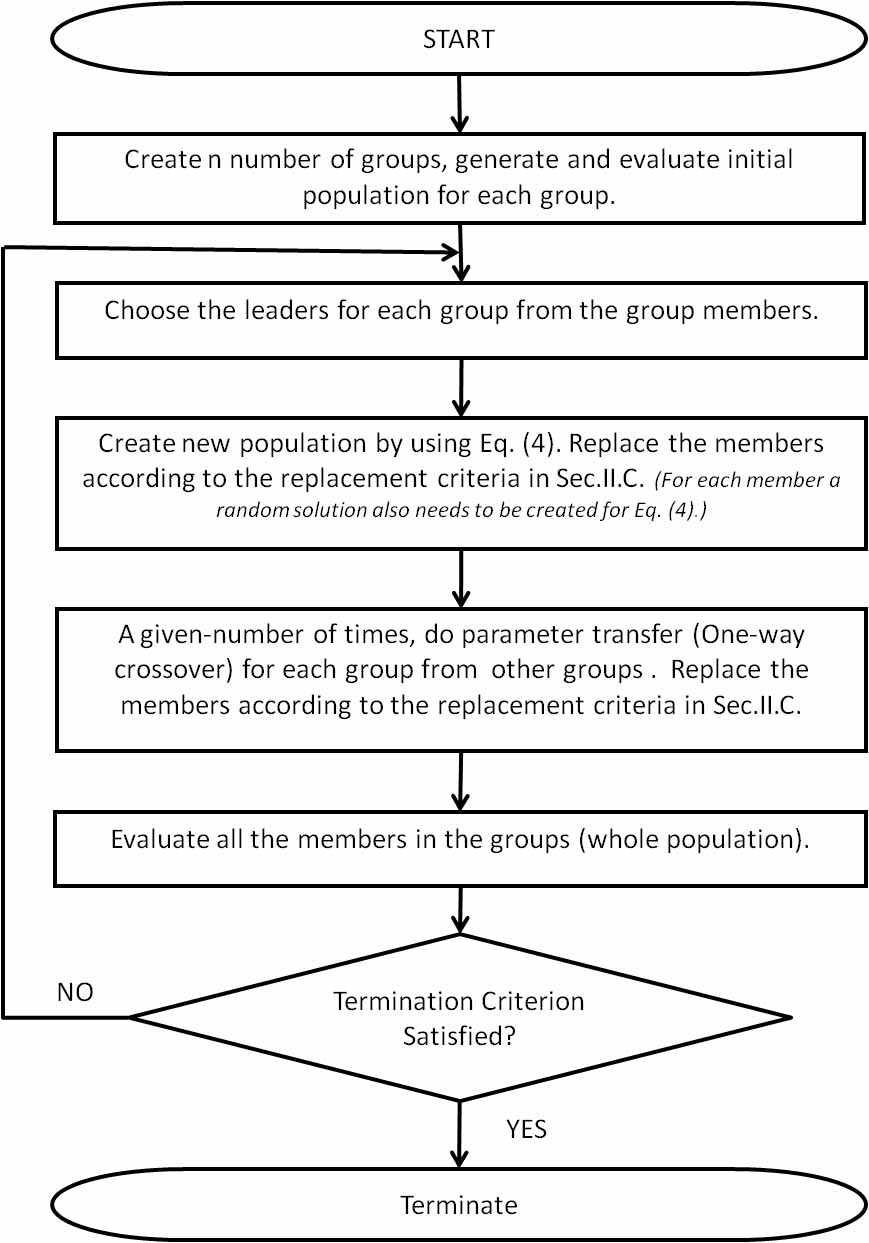}
\caption{The flow chart of the group leaders optimization algorithm}
\label{fig:gloa}
\end{figure*}

The flow chart of the algorithm for our optimization problem is drawn in Fig.\ref{fig:gloa}. 
For more information about the algorithm, the reader is referred to Ref.\cite{Daskin}.

\subsection{Parameters in the Optimization}

The parameters for the algorithm and $max_{gates}$ are defined in the previous subsections,
 the rest of the parameters used in the optimization are defined as follows:
 The default gate set consists of
 the rotation gates ($R_x, R_y, R_z, R_{zz} $); $X, Y, Z$ which are the Pauli operators $\sigma_x, \sigma_y,$ and $ \sigma_z$, respectively; the square root of $X$ gate $(V)$; 
the complex conjugate of $V$  gate ($V^\dagger$), $S$ gate, $T$ gate, and $H$ (Hadamard) gate; and the controlled versions of these gates. For the matrix representation of these gates, please see Appendix. The angle values for the rotation gates are taken as in the range  $[0,2\pi]$.

The test cases are divided into two categories: the known quantum algorithms  
and the unitary propagators of the molecular Hamiltonians. In the case of quantum algorithms the number of iterations is limited to 6000  iterations for the four-qubit quantum Fourier transform and 2000 for the rest while in  the case of molecular Hamiltonians it is limited to 15000. For both cases, the parameters of the algorithm: the number of group  is set to 25 and the number of population in each group is set to 15; so the total initial population is 375. The next two sections give and discuss the results.

\section{Circuit Designs for the Quantum Algorithms}
\label{Sec:Algorithms}
\begin{figure}[bp]
\centering
\includegraphics[width=3in]{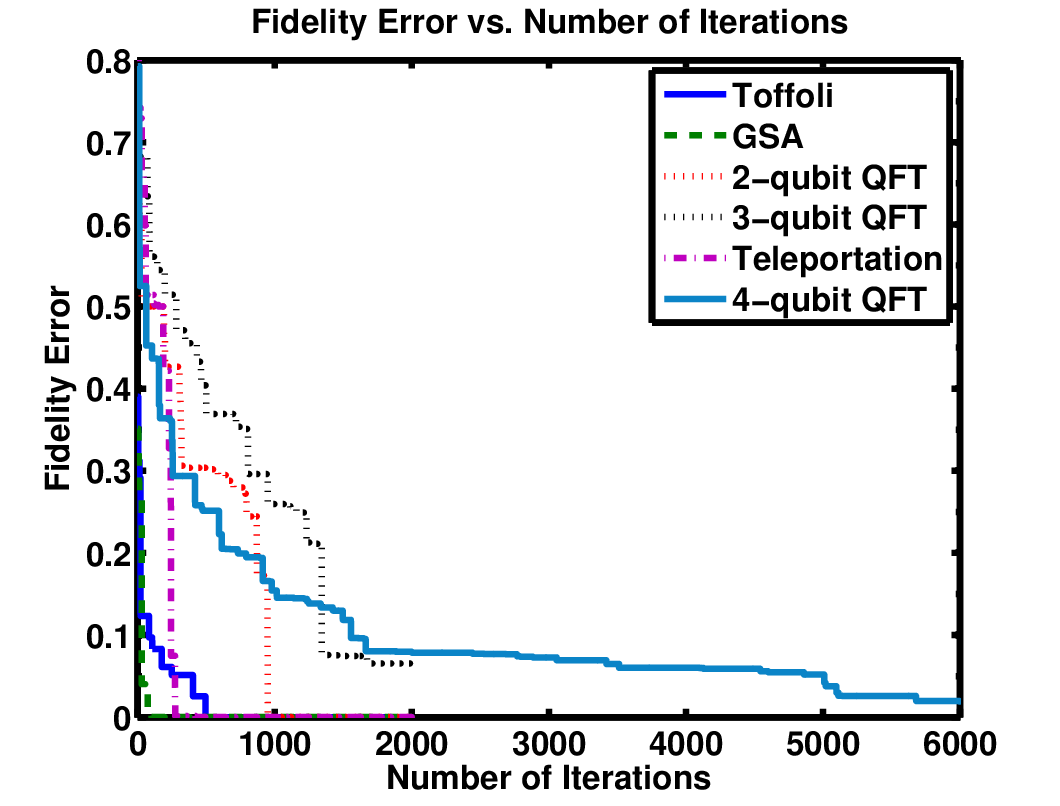}
\caption{The evolution of the fidelity error for quantum algorithms.}
\label{fig:error}
\end{figure}

The circuit designs for the cases of the known algorithms are not only important to find different circuit designs which may ease the implementation difficulties
 in different quantum computer models, 
but also to test the correctness, the efficiency, and the reliability of the optimization method on known results 
before using it to find the circuit representations of the more complex cases and the cases where the characteristics of the solutions are unknown. 
Hence, we use the optimization method to measure the ability of the method by finding circuit designs for the Toffoli gate,the Grover search algorithm, 
the quantum Fourier transform, and the quantum teleportation. The more details about these algorithms can be found in Ref.\cite{citeulike:541803}.
The resulting circuit designs with the descriptions of the problems are given in the following subsections. For each case, the evolution of the minimum fidelity error with respect to
  the number of iterations is plotted in Fig.\ref{fig:error}.

\subsection{The Toffoli Gate}
The Toffoli gate has two control, the first and the second, qubits  and one target, the third, qubit (see Appendix for the matrix representation). The circuit diagram for the unitary matrix 
of this gate shown in 
Fig.\ref{fig:toffoli}  which has the same length as the known circuit designs (see Ref.\cite{DiVincenzo01081998}). The algorithm reaches the exact solution in 500 iterations as shown in Fig.\ref{fig:error}.

\subsection{The amplification part of the Grover search algorithm}
\begin{figure}%
\centering
\includegraphics[width=2.5in]{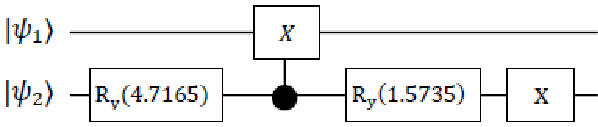}%
\caption{The circuit design for the two-qubit amplification part of the Grover search algorithm.}
\label{fig:grover}
\end{figure}

Grover's  algorithm searches an unstructured N-element list in $O(\sqrt{N})$ time compared to its classical counterpart which is $O(N)$.
The unitary operator for the amplification part of the Grover search algorithm (GSA) is constructed as \cite{Grover}:
\begin{equation}
\label{groveroperator}
    D_{ij}={\bigg\{} \begin{array}{cc}
					\frac{2}{N},& if\ i\ne j\\
					-1 + \frac{2}{N},& if \ i=j
					\end{array}.
\end{equation}
The algorithm reaches the explicit circuit diagram for the operator D for two qubits shown in Fig.\ref{fig:grover}
 after 100 iterations which is shown in Fig.\ref{fig:error}.

\subsection{The  Quantum Fourier Transform}
The quantum Fourier transform (QFT) is one of the key ingredients of the quantum factoring algorithm and many other quantum algorithms. 
The unitary operator of the QFT is constructed as \cite{citeulike:541803}:
\begin{equation}
 \frac{1}{\sqrt{2^n}}\left(
\begin{array}{cccccc}
1&1&1&\cdots&1\\
1&w&w^2&\cdots&1\\ 
1&w^2&w^4&\cdots&w^{\beta}\\
1&w^3&w^6&\cdots&w^{2\beta}\\
\vdots&\vdots&\vdots&\ddots&\vdots\\
1&w^{\beta}&w^{2\beta}&\cdots&w^{\beta^2}\\ 
\end{array}\right),
\end{equation}
where $\beta=2^n-1$ and $w=e^{2\pi i/2^n}$. 

\begin{figure}[h]%
\centering
\includegraphics[width=2.4in]{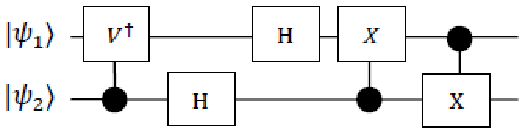}%
\caption{The circuit design for the two-qubit quantum Fourier transform.}
\label{fig:qft2}
\end{figure}

\begin{figure*}[htbp]
\centering
\includegraphics[width=4in]{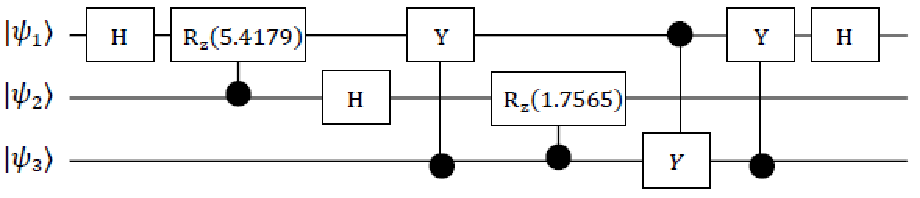}
\caption{The circuit design for the three-qubit quantum Fourier transform.}
\label{fig:qft3}
\end{figure*}

\begin{figure*}[htbp]
\centering
\includegraphics[width=5.5in]{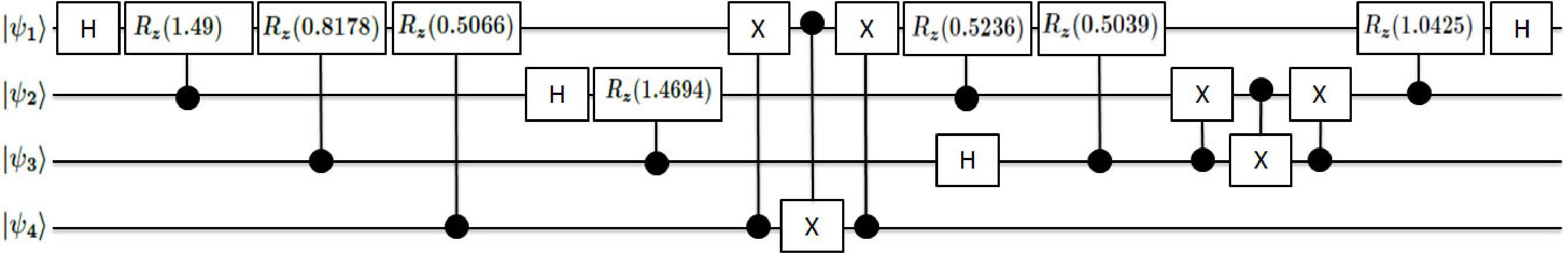}
\caption{The circuit design for the four-qubit quantum Fourier transform.}
\label{fig:qft4}
\end{figure*}

In our optimization,  we use the two-, three-, and four-qubit quantum Fourier transforms for which the circuit designs are found as in Fig.\ref{fig:qft2},
 Fig.\ref{fig:qft3}, and Fig.\ref{fig:qft4} 
respectively. The approximated circuit for the three-qubit case consists of 8 gates: 5 control and 3 single gates, while the circuit design for the same case in Ref.\cite{citeulike:541803}
 consist 6 control gates and requires swap operations at the end of the circuit. In Fig.\ref{fig:qft4}   13 control and 4 single gates form the approximated circuit for the four-qubit QFT in which there is only one more control gate in comparison to the general  circuit design of the QFT given in Ref.\cite{citeulike:541803} (the swap gates are also included in the comparison).   While the result for the two-qubit case is exact, for the three-and the four-qubit cases  small fidelity errors exist as shown in Fig.\ref{fig:error} because of the approximation of the angle values for the two rotation gates in the circuit.  

\subsection{The Quantum Teleportation}
Suppose Alice who has the first qubit wants to send information to Bob who has the second qubit. 
Quantum teleportation is a protocol which allows Alice to communicate an unknown quantum state of a 
qubit by using two classical bits in 
a way that Bob is able to reproduce the exact original state from these two classical bits 
\cite{1206629,citeulike:541803}.
Here, only the sender part of the quantum teleportation is used in the optimization since the unitary operator of the receiver part of the algorithm 
is similar to the sender part.  The matrix 
representation of the sender part of the algorithm is given in Appendix. Fig.\ref{fig:tlprt} shows the resulting circuit design for this matrix  which is found in 300 iterations (see Fig.\ref{fig:error}).  The circuit design in Fig.\ref{fig:tlprt} consists of 4 gates and has the same length as the known most efficient circuit designs given in  Ref.\cite{Williams99automateddesign, Brassard98teleportation}. 
\begin{figure}[h]%
\centering
\includegraphics[width=2.5in]{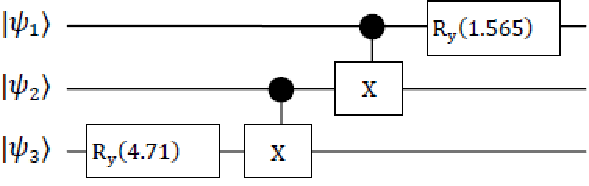}
\caption{The circuit design for the sender part of the quantum telportation.}
\label{fig:tlprt}
\end{figure}
\section{Circuit Designs for the Simulation of Molecular Hamiltonians}
\label{Sec:Hamiltonians}
Finding the low cost circuit representations of complex exponentials of the molecular Hamiltonians are important to be able to perform simulations on quantum computers. 
Here, after explaining the electronic Hamiltonian in the second quantized form and how to map the fermionic quantum operators to the standard quantum operators, we show how to use the optimization method to find quantum circuits for the molecular Hamiltonians and the simulation results for the water and the hydrogen molecules within the phase estimation algorithm.

Fermion model of quantum computation is defined through the spinless fermionic 
annihilation ($a_j$) and creation ($a_{j}^\dagger$) operators
 for each qubit j (j=1, ... , n), where the algebra of 2n elements obey the fermionic 
anti-commutation rules \cite{Ortiz}:
\begin{equation}
\{a_i,a_j \}=0,\  \{a_i, a_{j}^\dagger\}=\delta_{ij},
\end{equation}
where $\{A,B\}=AB+BA$ defines the anti-commutator. Using the Jordan-Wigner transformation \cite{Batista}, 
the fermion operators are mapped to the standard 
quantum computation operators through the Pauli spin operators:
\begin{equation}
\label{annihilationcreation}
\begin{split}
a_j\rightarrow \left(\prod_{k=1}^{j-1}{-\sigma_{z}^{k}}\right)\sigma_{-}^{j}=\left(-1\right)^{j-1}\sigma_{z}^{1}\sigma_{z}^{2} .... \sigma_{z}^{j-1}\sigma_{-}^{j}\\
a_j^\dagger\rightarrow \left(\prod_{k=1}^{j-1}{-\sigma_{z}^{k}}\right)\sigma_{+}^{j}=\left(-1\right)^{j-1}\sigma_{z}^{1}\sigma_{z}^{2} .... \sigma_{z}^{j-1}\sigma_{+}^{j}.
\end{split}
\end{equation}
 Once the electronic Hamiltonian is defined in second quantized form, the state space  
can be mapped to qubits. The molecular electronic Hamiltonian, in the Born-Oppenheimer approximation, is described in the second quantization form as \cite{Whitfield,Lanyon,Ovrum}:
\begin{equation}
\label{h2hamiltonian}
\mathcal{H}=\sum_{pq}{h_{pq}a_{p}^{\dagger}a_q}+\frac{1}{2}\sum_{pqrs}{h_{pqrs}a_{p}^\dagger a_{q}^\dagger a_{s}a_{r}},
\end{equation}
where the matrix elements $h_{pq}$ and $h_{pqrs}$ are the set of one- and
two-electron integrals.
Let the set of single-particle spatial functions constitute the molecular
 orbitals $\{\varphi(\textbf{r})\}^{M}_{k=1}$ and the set of spin orbitals $\{\chi(\textbf{x})\}_{p=1}^{2M}$ be defined with
$\chi_p=\varphi_i\sigma_i$ and the set of space-spin coordinates
$\textbf{x} = (\textbf{r}, \omega)$ where $\sigma_i$ is a spin function. The one-electron integral is defined as\cite{Whitfield}:
\begin{equation}
\begin{split}
h_{pq} &=\int{d\textbf{x} \chi_{p}^{*}(\textbf{x})\left(-\frac{1}{2}\triangledown^2-\sum_\alpha{\frac{Z_\alpha}{r_{\alpha \textbf{x}}}}\right)\chi_{q}(\textbf{x})}\\
&=\langle\varphi_p\mid H^{(1)}\mid\varphi_q\rangle\delta_{\sigma_p\sigma_q}
\end{split}
\end{equation}
and the two electron integral is:
\begin{equation}
\begin{split}
 h_{pqrs}&=\int{d\textbf{x}_1d\textbf{x}_2\frac{\chi_{p}^*(\textbf{x}_1)\chi_{q}^*(\textbf{x}_2)\chi_{s}(\textbf{x}_1)\chi_{r}(\textbf{x}_2)}{r_{12}}}\\
&=\langle\varphi_p\mid\langle\varphi_q\mid H^{(2)}\mid\varphi_r\rangle\mid\varphi_s\rangle\delta_{\sigma_p\sigma_q}\delta_{\sigma_r\sigma_s},
\end{split}
\end{equation}
where $r_{\alpha \textbf{x}}$ is the distance between the $\alpha^{th}$ nucleus and the electron, $r_{12}$ 
is the distance between electrons, $\triangledown^2$ 
is the Laplacian of the electron spatial coordinates, and $\chi_p(\textbf{x})$ is a selected single-particle basis:
 $\chi_p=\varphi_p\sigma_p$, $\chi_q=\varphi_q\sigma_q$, $\chi_r=\varphi_r\sigma_r$, and $\chi_s=\varphi_s\sigma_s$.
For detailed description of quantum computation for molecular energy simulations, see Withfield et al.\cite{Whitfield}.

\subsection{Phase Estimation Algorithm}
\begin{figure}[h]
\centering
\includegraphics[width=2.5in]{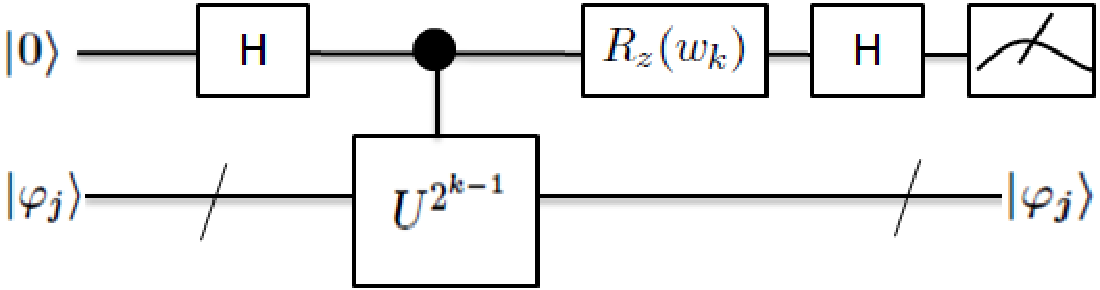}
\caption{The iterative phase estimation algorithm for the $k$th iteration\cite{miroslav,Abrams,Kitaev}. In the circuit $|\varphi_j\rangle$ is an eigenvalue of $U$, and the angle $w_k$ of the $R_z$ gate depends on the previous measured bits defined as $w_k=-2\pi(0.0\phi_k\phi_{k-1}...\phi_m)_{binary}$, where m is  the number of digits determining the accuracy of the phase $\phi_j$.  Note that $w_k$ is zero at the first iteration of the algorithm.}
\label{fig:ipea}
\end{figure}
Quantum computing provides an efficient method, the phase estimation algorithm (PEA)\cite{miroslav,Abrams,Kitaev}, to estimate the energy eigenvalues of a molecular Hamiltonian \cite{Williams}:
Suppose we have the unitary operator $U=e^{-i\mathcal{H}t/\hbar}$ for a Hamiltonian $\mathcal{H}$ with  energy eigenstates $|\psi_j\rangle$ corresponding energy eigenvalues $E_j$, i.e., 
$\mathcal{H}|\psi_j\rangle=E_j|\psi_j\rangle$. Since $E_j$ is an eigenvalue of $\mathcal{H}$; if $t$ and $\hbar$ are set to $1$, then $e^{-iE_j}$ is the eigenvalue of the unitary operator $U$.  Therefore, $N \times N$ unitary transformation $U$ has an orthonormal basis of eigenvectors $|\varphi_1\rangle$, $|\varphi_2\rangle$, ..., $|\varphi_N\rangle$ with eigenvalues $\lambda_j=e^{2\pi i \phi_j}$. The iterative PEA depicted  in Fig.\ref{fig:ipea} can be used to estimate the value of the phase $\phi_j$ which also allows us to determine the corresponding eigenvalue $E_j$ of the Hamiltonian $\mathcal{H}$.
The phase $\phi_j$ is obtained from the measurement results described as a binary expansion:
\begin{equation}
\begin{split}
\phi_j=&(0.\phi_1\phi_2...\phi_m)_{binary}
\\=&\phi_12^{-1}+\phi_22^{-2}...\ \phi_{m-1}2^{-m+1}+\phi_m2^{-m}.
\end{split}
\end{equation}

To find  the circuit equivalence of $U^{2^k}$ in Fig.\ref{fig:ipea}  the angle values of the rotation gates in the circuit represented by $U$ are multiplied by $2^k$ in each $k$th iteration of the algorithm  since $R_x(\theta)^{2^k}=R_x(2^k\theta)$, $R_y(\theta)^{2^k}=R_y(2^k\theta)$, and $R_z(\theta)^{2^k}=R_z(2^k\theta)$.

\subsection{Simulation of the Hydrogen Molecule}
The key challenge of exact quantum chemistry calculations is
the exponential growth of our description of the wave function
with the number of atoms. Consider a simple molecule like
methanol. Using only the 6-31G**  basis for the valence electrons,
there are 50 orbitals. The 18 valence electrons can be distributed in these orbitals in any way that satisfies the Pauli exclusion principle. This leads to about $10^{17}$ possible configurations making an exact or Full Configuration Interaction (FCI) calculation impossible.  Recently, a quantum algorithm for the solution of the FCI problem in polynomial time was proposed by Aspuru-Guzik et al. \cite{Alan-Science}. This algorithm employed the Hartree-Fock wave function as a reference for further treatment of the correlation effects by the FCI Hamiltonian on the quantum computer. 
Using an optical quantum computer, Lanyon et al. \cite{Lanyon}.
presented an experimental realization of quantum simulation of the energy spectrum of the hydrogen molecule. The key limitation in it is the representation of the simulated system's propagator.

One spatial function is needed per atom denoted $\varphi
_{H1}$ and $\varphi
_{H2}$  to describe the hydrogen molecule in minimal basis which is the minimum number of spatial functions required to describe the system. 
The molecular spatial-orbitals are defined by symmetry: 
$\varphi_{g}=\varphi
_{H1}+\varphi_{H2}$ and $\varphi_{u}=\varphi
_{H1}-\varphi_{H2}$; which correspond to four spinorbitals:
$|\chi_1\rangle=|\varphi_g\rangle|\alpha\rangle, |\chi_2\rangle=|\varphi_g\rangle|\beta\rangle, |\chi_3\rangle=|\varphi_u\rangle|\alpha\rangle,$ and 
$|\chi_4\rangle=|\varphi_u\rangle|\beta\rangle$. The STO-3G basis is used to evaluate the spatial integrals of the Hamiltonian which is defined as
$\mathcal{H}=H^{(1)}+ H^{(2)}$, where since $h_{pqrs}=h_{pqsr}$, $H^{(1)}$ and $ H^{(2)}$ are simplified as \cite{Lanyon,Whitfield,alan2}:
\begin{widetext}
\begin{equation}
\label{hamiltonianpart1}
H^{(1)}=h_{11}a_{1}^\dagger a_1+h_{22}a_{2}^\dagger a_2+h_{33}a_{3}^\dagger a_3+h_{44}a_{4}^\dagger a_4,
\end{equation}
and
\begin{equation}
\label{hamiltonianpart2}
\begin{split}
H^{(2)}&=h_{1221}a_{1}^\dagger a_{2}^\dagger a_{2} a_{1} +h_{3443}a_{3}^\dagger a_{4}^\dagger a_{4} a_{3} +h_{1441}a_{1}^\dagger a_{4}^\dagger a_{4} a_{1} + h_{2332}a_{2}^\dagger a_{3}^\dagger a_{3} a_{2}  +(h_{1331}-h_{1313})a_{1}^\dagger a_{3}^\dagger a_{3} a_{1}\\ &+(h_{2442}-h_{2424})a_{2}^\dagger a_{4}^\dagger a_{4} a_{2}+(h_{1423})(a_{1}^\dagger a_{4}^\dagger a_{2} a_{3} +a_{3}^\dagger a_{2}^\dagger a_{4} a_{1} )+(h_{1243})(a_{1}^\dagger a_{2}^\dagger a_{4} a_{3} +a_{3}^\dagger a_{4}^\dagger a_{2} a_{1} ).
\end{split}
\end{equation}
\end{widetext}
 Using the findings in ref.\cite{Whitfield} for the spatial integral values evaluated for atomic distance $1.401 a.u.$ in Eq.(\ref{hamiltonianpart1}) and Eq.(\ref{hamiltonianpart2}),
 the Hamiltonian matrix found as a matrix of order 16 (see the Appendix for the Hamiltonian matrix), so 4 qubits are required to implement the unitary propagator of this Hamiltonian which is found from $e^{-i\mathcal{H}t}$ (see the note \footnote{The time t in the equation is taken as 1. For the matrix exponentiation, we used the MATLAB
function \textit{expm} which uses the Pade approximation
with scaling and squaring \cite{Higham05thescaling}}). 

\begin{figure*}[htbp]
\centering
\includegraphics[width=5in]{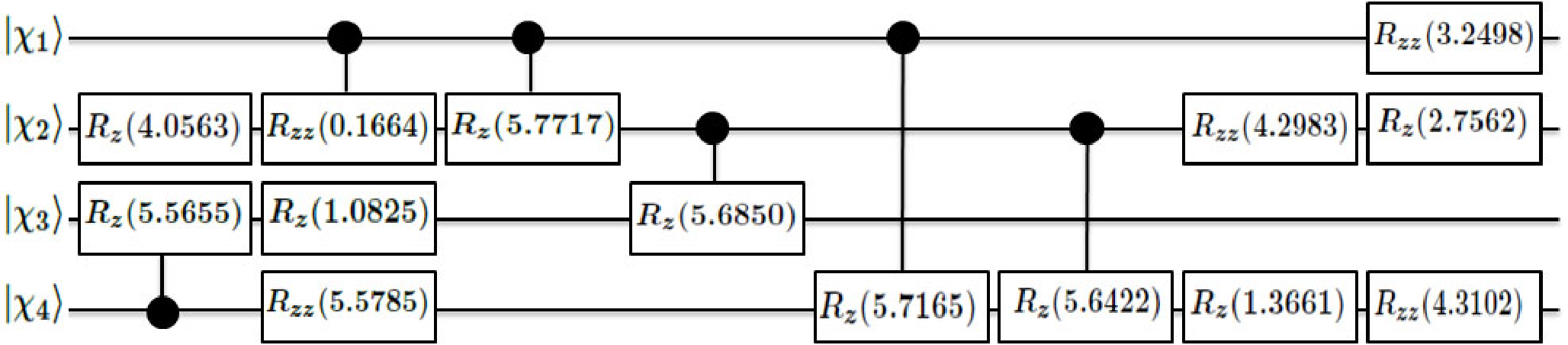}
\caption{The circuit design for the unitary propagator 
of the Hamiltonian of hydrogen molecule. The unitary propagator is found by using the spatial 
integral values in \cite{Whitfield} and
 the definitions for the annihilation and creation operators in 
Eq.(\ref{annihilationcreation}) into Eq.(\ref{hamiltonianpart1}) and Eq.(\ref{hamiltonianpart2}).}
\label{fig:hydrogeny}
\end{figure*}

\begin{figure*}[htbp]
\centering
\subfloat[The evolution of the fidelity error.]
{\includegraphics[width=3in]{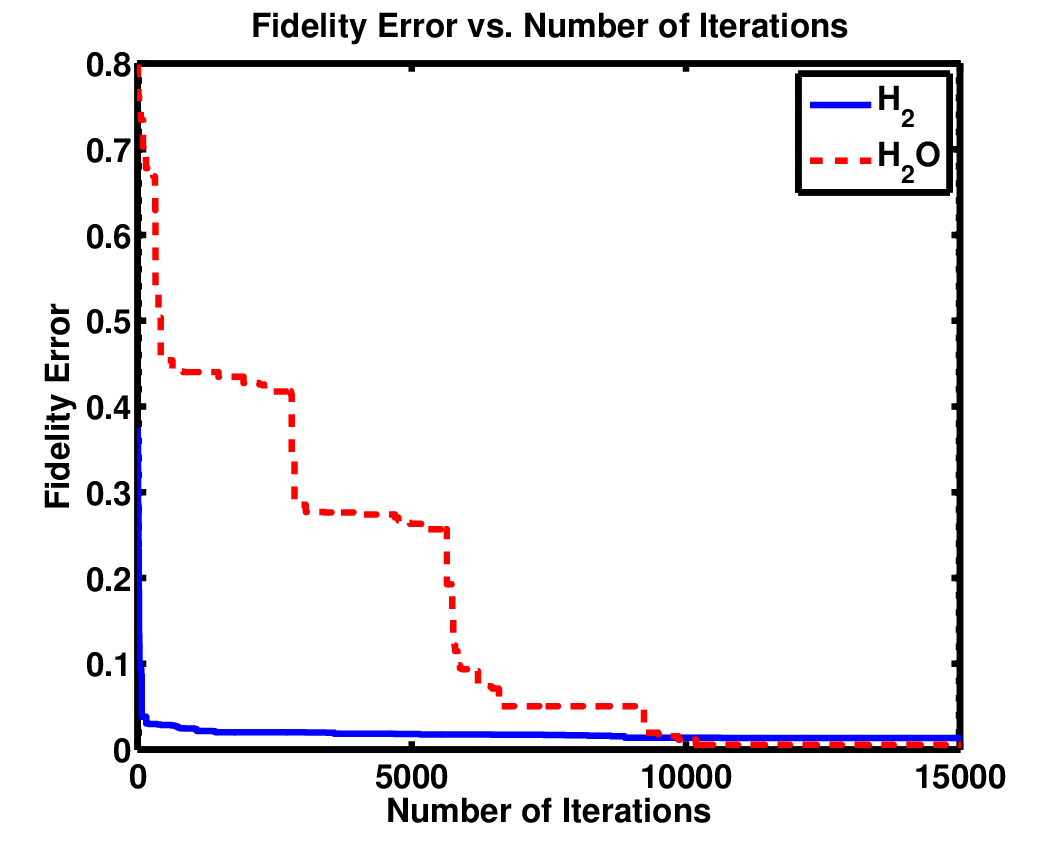}
\label{fig:errorH}}
\subfloat[The evolution of the cost.]{
\includegraphics[width=3in]{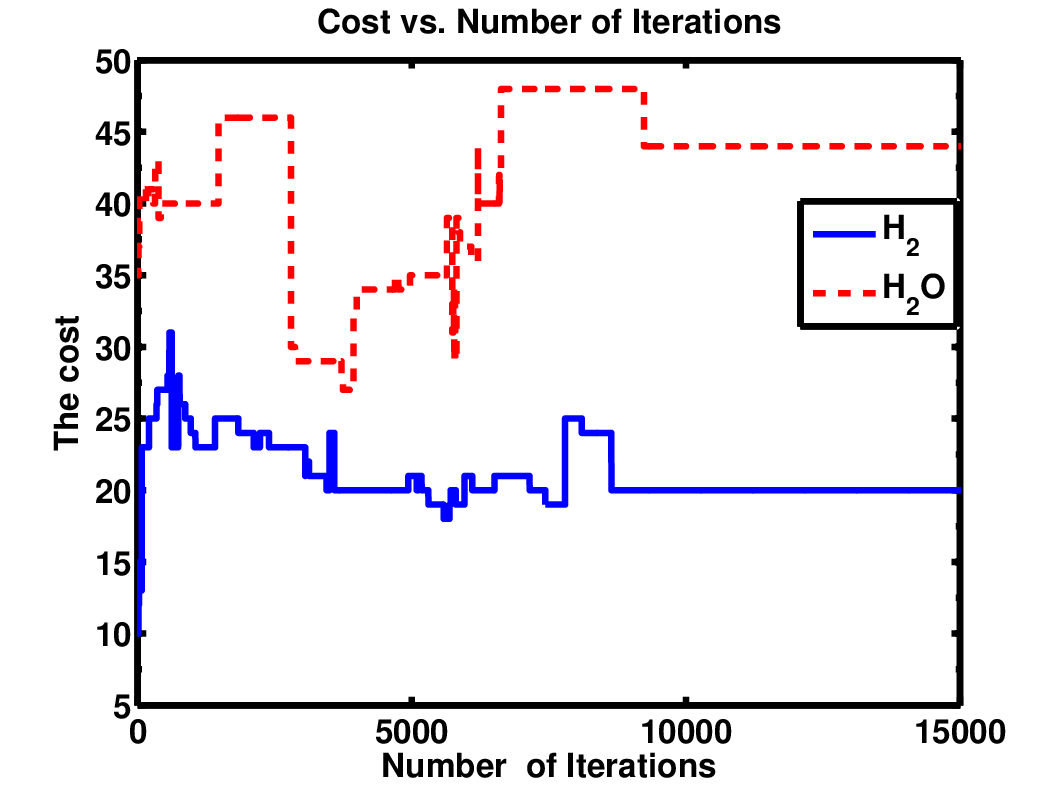}
\label{fig:costH}}
\caption{The evolutions of the cost and the error in the optimization for the exponentials of the Hamiltonians of the water and the hydrogen molecules.}
\label{fig:errorandcost}
\end{figure*}

 The decomposed circuit design for the unitary propagator of the hydrogen molecule is shown in Fig.\ref{fig:hydrogeny}. 
The circuit in Fig.\ref{fig:hydrogeny}
does not include the approximation error coming from the Trotter-Suzuki decomposition; however, it has some small errors 
which can be gauged from Fig.\ref{fig:errorH} showing the evolution of the error through the iterations of the optimization.  Therefore, the evolution of the cost is given in Fig.\ref{fig:costH}. 

The global phase with $e^{1.5i}$ is added to the beginning of the circuit in Fig.\ref{fig:hydrogeny} which allows the phase estimation algorithm to generate more accurate results. This global phase is estimated as: 
$phase=e^{sign(Im(p))acos(Re(p))i}$, where $p=Tr(U_aU_{H_2})/16$ and $U_a$ is the matrix representation of the found circuit.  The circuit including also the global phase  is simulated within the  phase estimation algorithm  (for each value the IPEA is run 20 times.). The phase and energy eigenvalues computed from the simulation are given in Table \ref{tab:eigenh2}  with the exact eigenvalues of the Hamiltonian matrix.

\begin{table}
\caption{The found and the corresponding exact  eigenvalues of the Hamiltonian of the hydrogen molecule}
\label{tab:eigenh2}
\begin{tabular}{ccc}
\hline
Phase& Found Energies& Exact Energies\\
\hline
0.0139	&	-0.0872	&	0.0000				\\
0.0314	&	-0.1971	&	0.2064					\\
0.0404	&	-0.2536	&	-0.2339					\\
0.0433	&	-0.2720	&	-0.3613				\\
0.0685	&	-0.4304	&	-0.3613					\\
0.0982	&	-0.6171	&	-0.4759					\\
0.1204	&	-0.7564	&	-0.4759					\\
0.1676	&	-1.0531	&	-0.8836					\\
0.1753	&	-1.1015	&	-1.1607					\\
0.1765	&	-1.1089	&	-1.1607					\\
0.1862	&	-1.1698	&	-1.2462					\\
0.2127	&	-1.3362	&	-1.2462					\\
0.2136	&	-1.3422	&	-1.2462					\\
0.2257	&	-1.4179	&	-1.2525					\\
0.2313	&	-1.4534	&	-1.2525					\\
0.2894	&	-1.8182	&	-1.8511					\\

\hline
\end{tabular}
\end{table}

\subsection{Simulation of  the Water Molecule}
The excited states of molecular systems are difficult to resolve by employing the
Hartree-Fock  wave function as an initial trial state. The main reason for
this difficulty is due to the fact that contributions from several
configuration state functions (CSF) must be considered if one
is seeking a reasonable overlap of the trial state with the exact
wave function.  Wang et al. \cite{Hefeng}  developed  a quantum algorithm to obtain the energy spectrum of molecular systems based on the
multiconfigurational self-consistent field (MCSCF) wave function. By using a MCSCF wave function as the initial guess, the excited states are accessible.
They demonstrate that such an algorithm can be used to obtain the energy spectrum of the water molecule.  The geometry used in the
calculation is near the equilibrium geometry (OH distance R = 1.8435 $a_0$ and the angle HOH = 110.57). With a complete active space type MCSCF method for the excited-state simulation, the CI space is composed of 18
CSFs, so 5 qubits are required to represent the wave function. 

\begin{figure*}[htbp]
\centering
\includegraphics[width=4in]{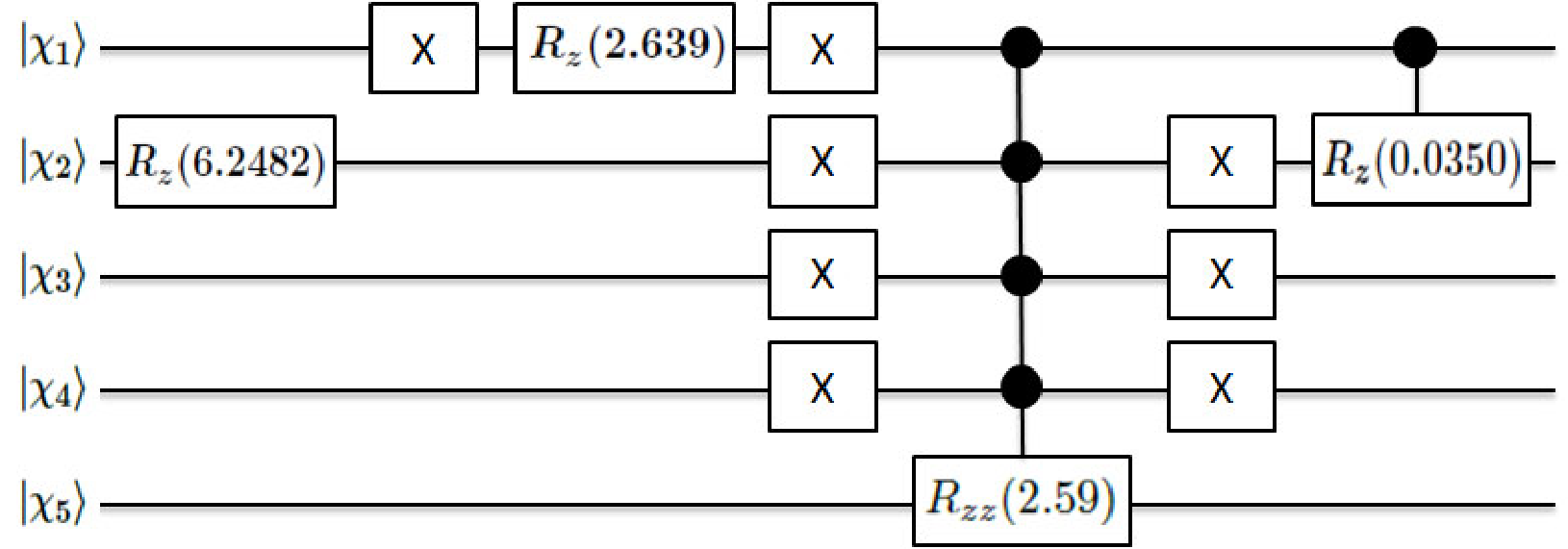}
\caption{The circuit design for the unitary propagator 
of the water molecule.}
\label{fig:h2o}
\end{figure*}

After finding the molecular Hamiltonian $\mathcal{H}$ as a matrix of order 18, 
we deploy the same idea as in Ref.\cite{Veis} and define the unitary operator as:
\begin{equation}
\hat{U}_{H_2O}=e^{i\tau (E_{max}-\mathcal{H})t}
\end{equation}
where $\tau$ is defined as:
\begin{equation}
\tau=\frac{2\pi}{E_{max}-E_{min}}.
\end{equation}
$E_{max}$ and $E_{min}$ are the expected maximum and minimum energies. The choice of $E_{max}$ and $E_{min}$ must cover all the eigenvalues of the Hamiltonian to obtain the correct results.  The final energy $E_j$  is found from the expression:

\begin{equation}
E_j=E_{max}-\frac{2\pi \phi_j}{\tau},
\end{equation}
where $\phi_j$ is the corresponding phase of the $E_j$. 
Since the eigenvalues of the Hamiltonian of the water molecule are between $-80 \pm \epsilon$ and $-84\pm \epsilon$ ($\epsilon \le0.1$), taking $E_{max}=0$ and $E_{min}=-200$ gives the following (see the note \footnotemark[\value{footnote}]):
\begin{equation}
\hat{U}=e^{\frac{-i2\pi H}{200}t}.
\end{equation} 
Fig.\ref{fig:h2o} shows the circuit diagram  for this unitary operator. The cost  of the circuit  is 44 (see Fig.\ref{fig:costH}) determined by summing up the cost of each gates in the circuit. The evolution of the fidelity error with respect to the number of iterations is plotted in 
Fig.\ref{fig:errorH}. 
Since we take $E_{max}$ as zero, this deployment does not require any extra quantum gate for the implementation within the phase estimation algorithm.
The simulation of this circuit within the iterative PEA results the phase and energy eigenvalues given in Table \ref{tab:eigenh2o}: the left two columns are respectively the computed phases and the corresponding energies, 
while the most right column of the matrix is the eigenvalues of the Hamiltonian of the water molecule (for each value of the phase, the IPEA is run 20 times).
\begin{table}
\caption{The found and the exact energy eigenvalues of the water molecule}
\label{tab:eigenh2o}
\begin{tabular}{ccc}
\hline
Phase& Found energy&  Exact energy \\
\hline
0.4200	&	-84.0019	&	-84.0021	\\		
0.4200	&	-84.0019	&	-83.4492	\\		
0.4200	&	-84.0019	&	-83.0273	\\		
0.4200	&	-84.0019	&	-82.9374	\\		
0.4200	&	-84.0019	&	-82.7719	\\		
0.4200	&	-84.0019	&	-82.6496	\\		
0.4200	&	-84.0019	&	-82.5252	\\		
0.4200	&	-84.0019	&	-82.4467	\\		
0.4144	&	-82.8884	&	-82.3966	\\		
0.4144	&	-82.8884	&	-82.2957	\\		
0.4144	&	-82.8884	&	-82.0644	\\		
0.4144	&	-82.8884	&	-81.9872	\\		
0.4144	&	-82.8884	&	-81.8593	\\		
0.4144	&	-82.8884	&	-81.6527	\\		
0.4144	&	-82.8884	&	-81.4592	\\		
0.4144	&	-82.8884	&	-81.0119	\\		
0.4122	&	-82.4423	&	-80.9065	\\		
0.4122	&	-82.4423	&	-80.6703	\\		
\hline
\end{tabular}
\end{table}
\section{Conclusion}
To be able to simulate Hamiltonians of atomic and molecular systems and  also apply quantum algorithms to solve different 
kinds of problems on quantum computers, it is necessary to find implementable quantum circuit 
designs including 
the minimum cost and number of quantum gate sequences. Since deterministic-efficient  quantum circuit design methodology is an open problem, 
we applied stochastic evolutionary optimization algorithm, GLOA,
 to search a quantum circuit design for the given unitary matrix representing a quantum algorithm or the 
unitary propagator of a molecular  Hamiltonian. 
In this paper, in addition to explaining the ways of the implementation and design of the 
optimization problem,  we give circuit designs for 
 the Grover search algorithm, the Toffoli gate, the quantum Fourier transform, and the 
quantum teleportation. Moreover, 
we find the circuit designs for the simulations of the water molecule and the hydrogen 
molecule by decomposing the unitary matrix operators found 
by following the fermionic model of quantum computation, and then simulate them within the phase estimation algorithm. 
In the case of the hydrogen molecule we found the number of gates needed to simulate the unitary operator 
 is 14 quantum gates (excluding the global phase) with the cost of 20. For the water molecule the cost of the number of operations  is found as 44 from the definition of the cost.
The approach is general and can be applied to generate the sequence of quantum gates for larger molecular systems.
Research is underway to generate the quantum circuit design for the simulation of the molecular Hamiltonian of $CH_2$ \cite{Veis}.
\section{Acknowledgments}
We would like  to thank  the NSF Center for Quantum Information and Computation for Chemistry, award number CHE-1037992, for financial support of this project. 
\bibliographystyle{apsrev4-1}
\bibliography{qpaper} 
\section{Appendix}
The matrix representation of quantum gates and algorithms used in the optimization as follows\cite{PhysRevA.52.3457,citeulike:541803,1206629}: 
\\
$X$, $Y$, and $Z$ gates which are 
the Pauli operators $\sigma_x$, $\sigma_y$, and $\sigma_z$ and Hadamard gate :
\begin{equation}
\begin{split}
X=& \left(\begin{array}{cc} 0&1\\1&0\\ \end{array}\right),\  \ \ \ Y =\left(\begin{array}{cc} 0&-i\\i&0\\ \end{array}\right),\\
Z =&\left(\begin{array}{cc} 1&0\\0&-1\\ \end{array}\right),\  H=\frac{1}{\sqrt{2}}\left(\begin{array}{cc} 1&1\\1&-1\\ \end{array}\right). 
\end{split}
\end{equation}

S gate and T, $\frac{\pi}{8}$, gate are:
\begin{equation}
S =\left(\begin{array}{cc} 1&0\\0&i\\ \end{array}\right),\ 
T=\left(\begin{array}{cc} 1&0\\0 &exp(i\frac{\pi}{4})\\ \end{array}\right).
\end{equation}

Square root of NOT (X) gate and its complex conjugate are:
\begin{equation}
V =\frac{1}{2}\left(\begin{array}{cc} 1 + i&1 - i\\1 - i&1 + i\\ \end{array}\right),\ 
V^\dagger=\frac{1}{2}\left(\begin{array}{cc} 1 - i&1+i\\1+i&1 - i\\ \end{array}\right) .
\end{equation}

Rotation gates are:
\begin{equation}
\begin{split}
R_x(\theta)&=\left(\begin{array}{cc}\cos(\frac{\theta}{2})& i\sin(\frac{\theta}{2})\\ i\sin(\frac{\theta}{2})&\cos(\frac{\theta}{2})\\ \end{array}\right),\\
R_y(\theta)&=\left(\begin{array}{cc}\cos(\frac{\theta}{2})&\sin(\frac{\theta}{2})\\-sin(\frac{\theta}{2})&\cos(\frac{\theta}{2})\\ \end{array}\right),\\
R_z(\theta)&=\left(\begin{array}{cc} 1&0\\0&exp(i\theta)\\ \end{array}\right),\\ 
R_{zz}(\theta)&=\left(\begin{array}{cc} exp(i\theta)&0\\0&exp(i\theta)\end{array}\right). 
\end{split}
\end{equation}
\\
The matrix representation of the Toffoli gate is as follows:
\begin{equation}
\left(\begin{array}{cccccccc}
   1&0&0&0&0&0&0&0\\0&1&0&0&0&0&0&0\\0&0&1&0&0&0&0&0\\
0&0&0&1&0&0&0&0\\0&0&0&0&1&0&0&0\\0&0&0&0&0&1&0&0\\0&0&0&0&0&0&0&1\\0&0&0&0&0&0&1&0
 \end{array}\right).
\end{equation}
\\
The matrix representation of the sender part of the quantum teleportation is as follows \cite{Yabuki00geneticalgorithms,reid2005}:
 \begin{equation*}
\frac{1}{2}\left(\begin{array}{cccccccc}
1&0&-1&0&0&1&0&1\\
0&1&0&-1&1&0&1&0\\
0&1&0&1&1&0&-1&0\\
1&0&-1&0&0&1&0&1\\
-1&0&-1&0&0&1&0&1\\
0&-1&0&1&1&0&1&0\\
0&-1&0&-1&1&0&-1&0\\
-1&0&-1&0&0&1&0&-1\\
                \end{array}\right).
\end{equation*}

The Hamiltonian of the hydrogen molecule which is found by using the spatial integral values in Ref.\cite{Whitfield} is as follows:
\vspace{2cm}
\begin{widetext}
\tiny
\begin{center}
$
\left(\begin{array}{l l l l l l l l l l l l l l l l l l l l l}
0.2064	&	0	&	0	&	0	&	0	&	0	&	0	&	0	&	0	&	0	&	0	&	0	&	0	&	0	&	0	&	0	&	\\
0	&	-1.1607	&	0	&	0	&	0	&	0	&	0	&	0	&	0	&	0	&	0	&	0	&	0	&	0	&	0	&	0	&	\\
0	&	0	&	-1.1607	&	0	&	0	&	0	&	0	&	0	&	0	&	0	&	0	&	0	&	0	&	0	&	0	&	0	&	\\
0	&	0	&	0	&	-1.8305	&	0	&	0	&	0	&	0	&	0	&	0	&	0	&	0	&	0.1813	&	0	&	0	&	0	&	\\
0	&	0	&	0	&	0	&	-0.3613	&	0	&	0	&	0	&	0	&	0	&	0	&	0	&	0	&	0	&	0	&	0	&	\\
0	&	0	&	0	&	0	&	0	&	-1.2462	&	0	&	0	&	0	&	0	&	0	&	0	&	0	&	0	&	0	&	0	&	\\
0	&	0	&	0	&	0	&	0	&	0	&	-1.0649	&	0	&	0	&	-0.1813	&	0	&	0	&	0	&	0	&	0	&	0	&	\\
0	&	0	&	0	&	0	&	0	&	0	&	0	&	-1.2525	&	0	&	0	&	0	&	0	&	0	&	0	&	0	&	0	&	\\
0	&	0	&	0	&	0	&	0	&	0	&	0	&	0	&	-0.3613	&	0	&	0	&	0	&	0	&	0	&	0	&	0	&	\\
0	&	0	&	0	&	0	&	0	&	0	&	-0.1813	&	0	&	0	&	-1.0649	&	0	&	0	&	0	&	0	&	0	&	0	&	\\
0	&	0	&	0	&	0	&	0	&	0	&	0	&	0	&	0	&	0	&	-1.2462	&	0	&	0	&	0	&	0	&	0	&	\\
0	&	0	&	0	&	0	&	0	&	0	&	0	&	0	&	0	&	0	&	0	&	-1.2525	&	0	&	0	&	0	&	0	&	\\
0	&	0	&	0	&	0.1813	&	0	&	0	&	0	&	0	&	0	&	0	&	0	&	0	&	-0.2545	&	0	&	0	&	0	&	\\
0	&	0	&	0	&	0	&	0	&	0	&	0	&	0	&	0	&	0	&	0	&	0	&	0	&	-0.4759	&	0	&	0	&	\\
0	&	0	&	0	&	0	&	0	&	0	&	0	&	0	&	0	&	0	&	0	&	0	&	0	&	0	&	-0.4759	&	0	&	\\
0	&	0	&	0	&	0	&	0	&	0	&	0	&	0	&	0	&	0	&	0	&	0	&	0	&	0	&	0	&	0	&	\\
\end{array}\right)$
\end{center}
\end{widetext}

\end{document}